\documentstyle[11pt,epsfig]{article}

\DeclareMathVersion{bold}

\newcommand{\pdf}{\emph{pdf}}
\newcommand{\etal}{\emph{et al.} }

\author{M. L\'opez-Caniego$^{1,2}$, D. Herranz$^{3}$,
  J.L. Sanz$^{1}$  and R.B. Barreiro$^{1}$  \\
  \\
  {\small $^{1}$ IFCA, Avda. de Los Castros s/n, 39905 Santander, Spain} \\
  {\small $^{2}$ Depto. de F\'\i sica Moderna, Univ. de Cantabria,} \\
  {\small Avda. de Los Castros, s/n, 39005 Santander, Spain} \\
  {\small $^{3}$ Istituto di Scienze e Tecnologie
    dell'Informazione ``A. Faedo'',} \\
  {\small CNR, via Moruzzi 1, 56124 Pisa, Italy}}

\title{Detection of point sources on two-dimensional
  images based on peaks}

\begin{document}

  \maketitle

  \abstract{This article considers the detection of point sources in
    two dimensional astronomical images. The detection scheme we
    propose is based on peak statistics. We discuss the example of the 
    detection of far galaxies in Cosmic Microwave Background
    experiments 
    throughout the paper, although  
    the method we present is totally general and can be used in
    many other fields of data analysis. 
    We assume
    sources with a Gaussian profile --that is a fair approximation of
    the profile of a point source convolved with the detector beam in
    microwave experiments-- 
    on a background modeled by a homogeneous and isotropic Gaussian
    random field characterized by a scale-free power spectrum. Point
    sources are enhanced with respect to the background by means of
    linear filters. After filtering, we identify local maxima
    and apply our
    detection scheme, a 
    Neyman-Pearson detector that defines our region of acceptance
    based on the  a priori \pdf  \ of the sources
    and the ratio of number densities.
    We study the different performances of some linear filters that
    have been used in this context in the literature: the Mexican Hat
    wavelet, the matched filter and the scale-adaptive filter. We
    consider as well an extension to two dimensions of the 
    biparametric scale adaptive 
    filter (BSAF). The BSAF
    depends on two parameters which are determined by maximizing the
    number density of real detections while 
    fixing the number density of spurious
    detections. For our detection criterion the BSAF 
    outperforms the other filters in the interesting case of white noise. \\
    \\
      {\bf keywords:} methods: analytical -- methods: data analysis --
      techniques: image processing}

  \section{INTRODUCTION} \label{sec:intro}

  A very challenging aspect of data analysis in astronomy
  is the detection of pointlike sources embedded in one
  and two dimensional images. 
  Some common examples are the separation of individual stars in 
  crowded optical images, the identification of emission and
  absorption lines in noisy one dimensional spectra 
  and the detection of faint extragalactic objects at microwave
  frequencies. 
  This latter case, for example, is one of the most critical issues
  for the new generation of experiments that observe the Cosmic
  Microwave Background (CMB).

  The CMB is the remnant of the radiation that filled the Universe
  immediately after the Big Bang. 
  This weak radiation can provide us with answers to one of the most important
  set of questions asked in modern science - how the Universe did
  begin, how it evolved to the state we observe today, and how it will
  continue to evolve in the future. 
  Unfortunately, we do not measure
  the CMB alone but a mixture of it with instrumental noise and other
  astrophysical radiations that are usually referred to as
  \emph{foregrounds}. 

  Some foregrounds are due to our own
  Galaxy, for example the thermal emission due to dust grains in the
  Galactic plane or the synchrotron emission by relativistic electrons
  moving along the Galactic magnetic field. These foregrounds appear
  as diffuse emission in the sky, and
  their spectral behaviours (the way the emission scales from one
  wavelength of observation to another) are reasonably well known. 
  Another foreground with a well known spectral behaviour is the
  Sunyaev-Zel'dovich effect, which 
  is due to the hot gas contained in galaxy clusters that distorts
  the energy distribution of CMB photons. 
  Foreground emissions carry information about the Galaxy
  structure, composition and physical parameters as well as about the
  number, distribution and evolution of galaxy clusters that map the
  distribution of matter in the
  Universe. Therefore, the study of the different foregrounds has great
  scientific relevance by itself. 
  In order to properly study the CMB and the different foregrounds it
  is mandatory to \emph{separate} the signals
  (\emph{components}) that are mixed in the observations. This can be
  done by
  observing the sky at a number of frequencies at least as
  big as the number of components and then applying some statistical
  \emph{component separation} method in order to recover the different
  astrophysical signals. Several component separation techniques
  have been suggested, including blind (Baccigalupi \etal 2000,
  Maino \etal 2002,
  Delabrouille \etal 2003),
  semi-blind (Bedini \etal 2004) and non-blind (Hobson \etal 1998,
  Bouchet and Gispert 1999, Stolyarov \etal 2002, Barreiro \etal 2004)
  approaches. 

  Another important foreground is due to
  the emission of far galaxies. Since the typical angular size of the
  galaxies in the sky is a few arcseconds and the angular resolution
  of the microwave detectors is typically greater than a few
  arcminutes\footnote{For example, the upcoming ESA's Planck
  satellite will
  have angular resolutions ranging from 5 arcminutes (for the
  217-857 GHz channels) to 33 arcminutes (for the 30 GHz channel).}, galaxies
  appear as points to the detector, which is unable to resolve their
  inner structure. Therefore, they are usually referred
  to as \emph{extragalactic point sources} (EPS) in the CMB jargon.
  Note that, however, they do not appear 
  as points in the images but as
  the convolution of a pointlike impulse with the angular response of
  the detector (``beam''). 
  The instruments (radiometers and bolometers) that are used in CMB
  experiments have angular responses that are approximately Gaussian
  and therefore EPS appear as small Gaussian (or nearly Gaussian)
  spots in the images\footnote{It is also common to speak of
  \emph{compact sources}, describing a source that is comparable to
  the size of the beam being used. Non-pointlike sources (such as
  large galaxy clusters with arcminute angular scales) will have
  more complicated responses when convolved with a beam, but if the
  source profile is known it is always possible to apply the methods
  presented in this work.}. 

  The problem with EPS is that galaxies are a very heterogeneous
  bundle of objects, from the radio galaxies that emit most of their
  radiation in the low frequency part of the electromagnetic spectrum
  to the dusty galaxies that emit mainly in the infrared (Toffolatti
  \etal 1998, Guiderdoni \etal 1998, Tucci \etal
  2004).   
  This makes it impossible to consider all of them as a single foreground
  to be separated from the other by means of multi-wavelength
  observations and statistical component separation techniques.
  EPS constitute an important contaminant in CMB studies
  at small angular scales (Toffolatti \etal 1998), affecting the
  determination of the CMB angular power spectrum and hampering the
  statistical study (e.g. the study of Gaussianity) of CMB and other
  foregrounds at such scales. Moreover, while there are good galaxy
  surveys at radio and infra-red frequencies, the microwave window of
  the electromagnetic spectrum is a practically unknown zone for
  extragalactic astronomy. Therefore, it is important to have 
  detection techniques that are able to detect EPS with
  fluxes as low as possible.

\begin{figure}
    \epsfxsize=75mm
    \includegraphics[width=14cm,angle=270]{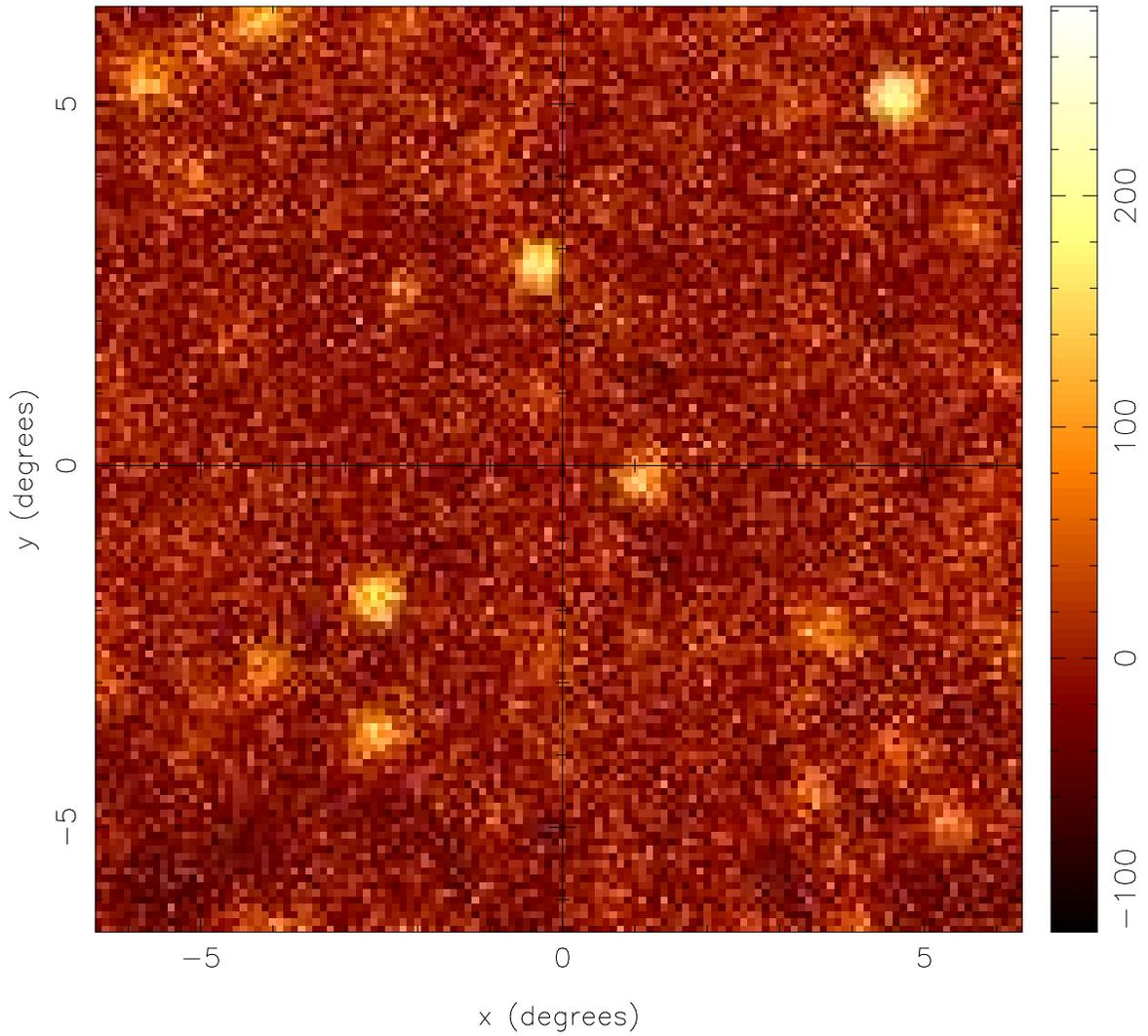}
    \caption{Residual map of a  $12.8 \times 12.8$ sqr deg sky patch at 30
      GHz after the application of a Maximum Entropy component
      separation. The residual map is obtained by subtracting from the 30
      GHz map the different components (CMB and foregrounds) given by the
      Maximum Entropy algorithm. Bright point sources appear as spots in
      the images whereas faint point sources are masked by the residual noise.}
    \label{fig:example_residual}
  \end{figure}

  One possibility is to consider the EPS emission at each frequency as
  an additional noise to be considered in the equations of a
  statistical component separation method. Once the 
  algorithm has separated the different components, the residual that
  is obtained by subtracting the output foregrounds from the original
  data should contain the EPS plus the instrumental noise and some
  amount of foreground residuals that remain due to a non-perfect
  separation. As an example, Figure~\ref{fig:example_residual} shows 
  the residual at 30 GHz after applying a Maximum Entropy component separation
  algorithm (Hobson \etal 1999) to a $12.8 \times 12.8$ sqr deg 
  simulated sky patch as would be observed by the Planck
  satellite. The brightest point sources can be clearly observed over
  the residual noise. However, fainter point sources are still
  masked by a residual noise that is approximately Gaussian and must
  be detected somehow. Besides, the situation is more complex because
  the presence of bright EPS in the data affects the performance of
  the component separation algorithms so the recovered components
  are contaminated by point sources in a way that is difficult to
  control. Therefore, any satisfactory method should detect and extract at
  least the bright sources before the component separation. Then,
  after separation some additional low intensity EPS could be 
  detected from the
  residual maps such as the one in Figure~\ref{fig:example_residual}.

  Several techniques based on linear filters have been proposed
  in the literature for the detection of point sources in CMB data.
  Linear filtering techniques are suitable for this problem because
  they can isolate structures with a given characteristic scale,
  as is the case of pointlike sources, while cancelling the
  contribution of diffuse foregrounds. Among the methods
  proposed in the literature, we emphasize the Mexican Hat Wavelet
  (MHW, Cay\'on \etal 2000, Vielva \etal 2001a,b,2003), the classic
  \emph{matched filter} (MF, Tegmark and de Oliveira-Costa 1998), 
  the Adaptive Top Hat Filter (Chiang \etal 2002) and the scale-adaptive
  filter (SAF, Sanz \etal 2001, Herranz \etal 2002). Moreover, 
  linear filters
  can be used in combination with statistical component separation
  techniques in order to produce a more accurate separation of the
  different foregrounds (Vielva \etal 2001b). 
 
  The goal of filtering is to enhance the contrast between the
  source to be detected and the background that masks it. 
  For example, if we filter the image in
  Figure~\ref{fig:example_residual}, assuming that the background can be
  characterised by a white noise, with the well known matched filter
  (see section~\ref{sec:MF}) at the scale of the 30 GHz detector
  beam (FWHM=33 arcminutes) the signal to noise ratio of the sources
  increases by more than $25 \%$. Therefore, a source whose signal to
  noise ratio was $\sim 3$ before filtering becomes a source with signal
  to noise ratio $\sim 4$ and will be easier to detect.

  After
  filtering, a \emph{detection rule} is applied to the data
  in order to decide whether the source is present or not. 
  The usual detection approach in astronomy is \emph{thresholding}:
  for any given candidate (for example a local peak in the data)
  a positive detection is considered if the candidate  
  has a signal to noise ratio greater than a certain threshold
  (in many astronomical applications, a typical value 
  of this threshold is $5 \sigma$). This naive approach works fine
  for bright sources, but weak sources can be easily missed. 
  
  More sophisticated detection schemes can use 
  additional information in order to improve the detection.
  If the detection is performed by means of the study of the
  statistics of maxima in the images, such information
  includes not only the amplitude of the maxima but also
  spatial information related with the source profile, 
  for example the derivatives of the intensity; in our approach
  we will consider the amplitude, the curvature and the 
  shear of the sources (the last two quantities are given by the 
  properties of the beam in the case of point sources) to 
  discriminate between maxima of the background and real 
  sources. Moreover, in some cases 
  \emph{a priori} information on the distribution
  of intensity of the sources is known. We will therefore
  use a 
  Neyman-Pearson detector that
  uses the three above mentioned elements of information (amplitude,
  curvature and shear) of the maxima as well as the \emph{a priori}
  probability distribution of the sources. This technique
  has been successfully tested in 
  images of one-dimensional fields (L\'opez-Caniego
  \etal 2004a,b). In this work we will generalise it to two 
  dimensions.
  
  The overview of this work is as follows: in section~\ref{sec:peaks} we
  describe the statistics of the peaks for a two-dimensional
  Gaussian background in the absence and presence of a source.
  In section~\ref{sec:detection} we introduce the detection problem,
  define the
  region of acceptance and derive our detector.
  In section~\ref{sec:filters} we briefly review some of the linear filters
  proposed in the literature. In section~\ref{sec:numerical} we
  describe a probability 
  distribution of sources that is of interest and compare the 
  performance of the filters, regarding our choice of detector. Finally, 
  in section~\ref{sec:conclusions} we summarise our results. 

\section{PEAK STATISTICS} \label{sec:peaks}

  In this section we will study the statistics of peaks for a
  two-dimensional Gaussian background in both the absence and presence
  of a source. We will focus on three quantities that define the 
  properties of the peaks: the intensity of the field, the curvature
  and the shear at the position of the peak. The first quantity gives
  the amplitude of the peak. The curvature and the shear give
  information about the spatial structure of the peak and are related
  to its sharpness and eccentricity, respectively. 
  
  \subsection{Background}

    Let us assume a two-dimensional (2D) background represented by a
    Gaussian random field 
    $\xi (\vec{x})$ with average value $\langle \xi (\vec{x})\rangle = 0$
    and power spectrum $P(q)$, 
    \begin{equation}
      \langle \xi
      (\vec{Q})\xi^* (\vec{Q'})\rangle = P(q)\delta_{D}^2
      (\vec{Q}-\vec{Q'}), \ \ \ q\equiv |\vec{Q}|,
    \end{equation} 
    \noindent where $\xi (\vec{Q})$ is the Fourier transform of
    $\xi (\vec{x})$\footnote{Throughout this paper we will use the
      following notation for the 
      Fourier transform: the same symbol will be used for the real
      space and the Fourier space versions of a given function. The
      argument of the function will specify in each case which is the
      space we are referring to. For instance, $f(q)$ will be the Fourier
      transform of the function $f(x)$.}
    and $\delta_D^2$ is the Dirac distribution in 2D.
    
    We are interested in the distribution of maxima of the background
    with respect to the three variables already mentioned: intensity,
    curvature and shear. Let us define the \emph{normalized field
    intensity} $\nu$, the \emph{normalized curvature} $\kappa$ and the
    \emph{normalized shear} $\epsilon$ as:
    \begin{equation} \label{eq:def_nu_kappa_epsilon}
      \nu \equiv \frac{\xi}{\sigma_0},\ \ \ 
      \kappa \equiv \frac{\lambda_{1} + \lambda_{2}}{\sigma_2},\ \ \ 
      \epsilon \equiv \frac{\lambda_1 - \lambda_2}{2 \sigma_2},
    \end{equation}
    \noindent 
    where $\nu \in (-\infty ,\infty )$, $\kappa \in
    [0,\infty)$,
    $\epsilon \in [0,\kappa/2 )$, 
    $\lambda_1$ and
    $\lambda_2$ are the eigenvalues of the negative Hessian matrix,
    and the $\sigma_n²$ are defined as 
    \begin{equation} \label{eq:sigman_generic}
      \sigma_n^2 \equiv \frac{1}{2 \pi}\int_0^{\infty} dq \ q^{1+2n} P(q).
    \end{equation}
    \noindent
    The moment $\sigma_0$ is equal to the dispersion of the field.

    The
    distribution of maxima of the background in one dimension (1D)
    with respect to the
    intensity and curvature (the shear is not defined in 1D) 
    was studied by Rice
    (1954). If we generalize it to 2D, including the shear,
    the expected number density of
    maxima per intervals  $(\vec{x}, \vec{x} + d\vec{x})$, $(\nu ,\nu +
    d\nu )$, $(\kappa, \kappa + d\kappa )$ and $(\epsilon ,\epsilon +
    d\epsilon)$ is given by  
    \begin{eqnarray} \label{eq:nbackground} 
      n_b(\nu ,\kappa, \epsilon )  = \frac{8 \sqrt{3} \tilde{n}_b}{\pi
	\sqrt{1-\rho^2}} \epsilon (\kappa^2 - 4\epsilon^2)
      e^{-\frac{1}{2}\nu^2 - 4\epsilon^2 -\frac{(\kappa - \rho
	  \nu)^2}{2(1 - \rho^2)}},  
    \end{eqnarray} 
    \noindent where $\tilde{n}_b$ is the the expected total 
    number density of
    maxima (i.e. number of maxima per unit area $d\vec{x}$)  
    \begin{equation}  \label{eq:nb_total}
      \tilde{n}_b \equiv \frac{1}{4\pi \sqrt{3} ~\theta^2_m },
    \end{equation}
    \noindent and $\rho$ and $\theta_m$ are defined as
    \begin{equation} \label{eq:thetas_rho}
      \theta_m \equiv \sqrt{2} \frac{\sigma_1}{\sigma_2},\ \ \ 
      \rho \equiv \frac {\sigma_1^2}{\sigma_0 \sigma_2} =
      \frac{\theta_m}{\theta_c},\ \ \  
      \theta_c \equiv \sqrt{2}\frac{\sigma_0}{\sigma_1}.
    \end{equation}
    \noindent In the previous equations
    $\theta_c$ and $\theta_m $ are the coherence scale of the field
    and maxima, respectively.
    The formula in equation (\ref{eq:nbackground})
    can be derived from previous works (Bond and Efstathiou
    1987, Barreiro \etal 1997). 

  \subsection{Background plus point source}

    To the previous 2D background we add a source with
    a known spatial profile $\tau (\vec{x})$ and an amplitude $A$, so
    that the intensity due to the source at a given position $\vec{x}_0$ is
    $\xi_s(\vec{x}) = A \tau(\vec{x}-\vec{x}_0)$. 
    For simplicity,
    we will consider a spherical 
    Gaussian profile given by 
    \begin{equation}
      \tau (x) = \exp ({- x^2/2R^2}), \ \ \  x \equiv |\vec{x}|, 
    \end {equation}
    \noindent where $R$ is the Gaussian width (in the case of point
    sources convolved with a Gaussian beam, $R$ is the beam width). 
    We could easily
    consider other functional profiles\footnote{For example, more complicated
    beams or sources that are not pointlike but have some resolved
    structure.} without any loss of generality.  
    The expected number
    density of maxima per intervals $(\vec{x}, \vec{x} + d\vec{x})$,
    $(\nu ,\nu + d\nu )$, $(\kappa ,\kappa + d\kappa )$ and $(\epsilon
    ,\epsilon + d\epsilon)$, given a source of amplitude $A$ in such
    spatial interval, is 
    \begin{equation}  \label{eq:nsource}
      n(\nu ,\kappa, \epsilon |\nu_s) =  \frac{8 \sqrt{3}\,\tilde{n}_b}{\pi
	\sqrt{1-\rho^2}} \epsilon(\kappa^2 - 4\epsilon^2)
      e^{-\frac{1}{2}(\nu - \nu_s)^2 - 4\epsilon^2 -\frac{(\kappa -
	  2\kappa_s - \rho(\nu - \nu_s))^2}{2(1 - \rho^2)}} 
    \end{equation} 
    \noindent where 
    $\nu_s = A/\sigma_0$ is
    the normalized amplitude of the source, $\kappa_s = -
    A\tau^{\prime \prime}/ \sigma_2$ is the normalized
    curvature of the source
    and $\tau^{\prime \prime}$ is the second derivative of the source
    profile $\tau$ with respect to $x$ at the position of the source.
    Note that in 
    equation (\ref{eq:nsource}) 
    we are taking into account that the shear of the
    source is zero since we are considering a spherical profile.
    It is useful to define a quantity $y_s$ that is related to
    the curvature of the source:
    \begin{equation} \label{eq:ys}
      y_s \equiv - \frac{\theta^2_m
	\tau^{\prime \prime} }{\rho}, \ \ \
      \kappa_s = \frac{\nu_s y_s}{2}. 
    \end{equation}

\section{THE DETECTION PROBLEM} \label{sec:detection}

  Equations (\ref{eq:nbackground}) and (\ref{eq:nsource})
  can be used to decide whether a source is present or not in a data
  set.
  The tool that allows us to
  decide whether a point source is
  present or not in the data is called a \emph{detector}.
  In this section we will
  describe the 
  Neyman Pearson detector (NPD). We will study 
  its performance in terms of two quantities: the number of
  true detections and the number of false (spurious) detections 
  that emerge from the detection process.
  Our approach fixes
  the number density of spurious detections
  and determines the number density
  of true detections in each case.

  \subsection{The region of acceptance}
 
    We consider a peak in the 2D data set characterized by the normalized 
    amplitude, curvature and  shear $(\nu,\kappa,\epsilon)$. The number 
    density of background maxima  $n_b(\nu,\kappa,\epsilon)$ represents 
    the null hypothesis $H_0$  
    that the peak is due to the background in the absence of source.
    Conversely, the local number density of maxima 
    $n(\nu,\kappa,\epsilon)$ represents the alternative hypothesis,
    that the peak is due to the source added to the background. The
    local number density of maxima $n(\nu,\kappa,\epsilon)$ can be
    calculated as
    \begin{equation} \label{eq:nsource_integrated}
      n (\nu ,\kappa,\epsilon)= \int_0^{\infty} d\nu_s p(\nu_s)n(\nu,
      \kappa, \epsilon |\nu_s ). 
    \end{equation}
    In the last equation we have used the \emph{a priori} probability
    $p(\nu_s )$, that gives the amplitude distribution
    of the sources.
    
    We can associate to any region  $\mathcal{R}_*(\nu,\kappa,\epsilon )$ 
    in the $(\nu, \kappa, \epsilon)$ 
    parameter space 
    two number densities $n_b^*$ and $n^*$ 
    \begin{equation} \label{eq:nbstar}
      n_b^* = \int_{\mathcal{R}_*}d\nu \,d\kappa\,d\epsilon \,n_b(\nu
      ,\kappa,\epsilon),  
    \end{equation}
    \begin{equation} \label{eq:nstar}
      n^* = \int_{\mathcal{R}_*}d\nu
      \,d\kappa\,d\epsilon\,n(\nu,\kappa,\epsilon). 
    \end{equation}
    \noindent where  $n_b^*$ is the expected number density of spurious 
    sources, i.e. due to the background, in the region  
    $\mathcal{R}_*(\nu,\kappa,\epsilon)$, whereas $n^*$ is the number 
    density of maxima expected in the same region of the 
    $(\nu,\kappa,\epsilon)$ space in the presence of a local source.
    The region $\mathcal{R}_*$ will be called the 
    \emph{region of acceptance}. 

    In order to define the region of acceptance $\mathcal{R}_*$ that 
    gives the highest number density of detections $n^*$ for 
    a given number density of spurious detections $n_b^*$, we assume 
    a Neyman-Pearson Detector (NPD) 
    using number densities instead of probabilities
    \begin{equation}
      L(\nu ,\kappa,\epsilon )\equiv \frac{n(\nu ,\kappa,\epsilon
	)}{n_b(\nu ,\kappa,\epsilon )}\geq L_*,  
      \label{lik_nus}
    \end{equation}
    \noindent where $L_*$ is a constant. The proof follows the same 
    approach as for the standard Neyman-Pearson detector. If  
    $L\geq L_*$ we decide that the signal is present, 
    whereas if $L< L_*$ we decide that the signal is absent. From this
    ratio $L\geq L_*$,
    we derive the region of acceptance, that is given by  
    the sufficient linear detector $\varphi$ (see Appendix)
    \begin{equation} \label{eq:toapp}
      \mathcal{R}_*:  \varphi (\nu ,\kappa )\geq \varphi_* ,
    \end{equation}
    where $\varphi_*$ is a constant and $\varphi(\nu,\kappa)$ is given by 
    \begin{equation} 
      \varphi (\nu ,\kappa )\equiv a\nu + b\kappa, \ \ a \equiv \frac{1
	- \rho y_s}{1 - \rho^2}, \ \ b \equiv  \frac{y_s - \rho }{1 -
	\rho^2}. 
      \label{eq:phi}
    \end{equation}
    We remark that the detector is independent of the shear $\epsilon$. 
    This is due to the fact that we are considering a source with a 
    spherical profile with shear $\epsilon_s =0$. 
    If the profile is not spherical, the detector may depend on the shear.

  \subsection{Spurious sources and real detections}

    Given a region of acceptance  $\mathcal{R}_*$, we can calculate the 
    number density of spurious sources and the number density of 
    detections as given by equations (\ref{eq:nbstar}) and (\ref{eq:nstar})
    \begin{equation} \label{eq:nbstar1}
      n_b^* = \frac{\sqrt{3}\tilde{n}_b}{\sqrt{2\pi}} \int_{0}^\infty
      d\kappa\,(\kappa^2 -1 +e^{-\kappa^2})e^{-\frac{\kappa^2}{2}}
      {\rm erfc}\left(M\right), \ \ \ M \equiv \frac{\varphi_* -
      y_s\kappa}{a \sqrt{2(1-\rho^2)}}. 
    \end{equation}
    \begin{equation} \label{eq:nstar1}
      n^* = \frac{\sqrt{3}\tilde{n}_b}{\sqrt{2\pi}} \int_{0}^\infty
      d\nu_s p(\nu_s) \int_{0}^\infty d\kappa (\kappa^2 - 1 +
      e^{-\kappa^2})e^{-\frac{1}{2}(\kappa - \nu_s y_s)^2}{\rm
      erfc}\left(Q\right), 
    \end{equation}
    \begin{equation}
      Q \equiv M + \nu_s \frac{\rho y_s - 1}{\sqrt{2(1-\rho^2)}}.
    \end{equation}
    Our approach is to fix the number density of spurious detections
    and then to determine the region of acceptance that gives the
    maximum number of true detections. This can be done by
    inverting the equation (\ref{eq:nbstar1}) to obtain 
    $\varphi_* =\varphi_* (n^*_b/n_b; \rho , y_s)$. Once
    $\varphi_*$ is known, we can calculate the number density of
    detections using equation (\ref{eq:nstar1}).

\section{THE FILTERS} \label{sec:filters}

  Detection can, in principle, be performed on the raw data, but in
  most cases it is 
  convenient to transform first the data in order to enhance the
  contrast between the distributions 
  $n_b(\nu ,\kappa, \epsilon )$ and 
  $n(\nu ,\kappa, \epsilon)$.
  Hopefully, such an enhancement will
  help the detector to give better results (namely, a higher
  number of true detections). 
  In this paper we will focus in the use of linear filters
  as a means to transform the data in such a way. Filters
  are suitable for this task because background fluctuations that have
  variation scales different from the source scale can be easily
  filtered out while preserving the sources. 
  Different filters will
  improve detection in different ways: this paper
  compares the performance of several filters.
  The filter that gives
  the highest number density of detections, 
  for a fixed number density of spurious sources,
  will be the 
  preferred filter among the considered filters.

  Let us consider a filter 
  $\Psi (\vec{x}; R, \vec{b})$, where $R$ and $\vec{b}$ define a 
  scaling and a translation respectively.
  Since the sources we are considering
  are spherically symmetric and we assume that the background is
  statistically homogeneous and isotropic, we will consider
  spherically symmetric filters, 
  \begin{equation}
    \Psi(\vec{x}; R, \vec{b}) \equiv \frac{1}{R^2} \psi \left(
    \frac{|\vec{x} - \vec{b}|}{R} \right) .
  \end{equation}
  If we filter our 
  background with  $\Psi (\vec{x}; R, \vec{b})$, the filtered field is
  \begin{equation} \label{eq:filtered_field}
    w(R, \vec{b}) = \int d\vec{x}\,\xi (\vec{x})\Psi (\vec{x}; R, \vec{b}).
  \end{equation}
  \noindent The filter is normalized such that the amplitude of the
  source is the same after filtering:
  \begin{equation}
    \int d\vec{x} \, \tau(\vec{x}) \Psi(\vec{x}; R, \vec{0}) = 1.
  \end{equation}
  \noindent
  For the filtered field equation
  (\ref{eq:sigman_generic}) becomes
  \begin{equation}
    \sigma_n^2 \equiv 2 \pi \int_0^{\infty} dq \ q^{1+2n} P(q)
    \psi^2(q). 
  \end{equation}
  \noindent 
  The values of $\rho$, $\theta_m$, $\theta_c$ and all the
  derived quantities 
  change accordingly. The curvature of the filtered source
  $\kappa_s$ can be obtained through equation (\ref{eq:ys}), taking
  into account that for the filtered source 
  \begin{equation} \label{eq:curvature_source}
    -\tau^{\prime \prime}_{\psi} = \pi \int_0^{\infty}
    dq \ q^3 \tau(q) \psi(q).
  \end{equation}
  \noindent
  Note that the function $\psi(q)$ will depend as well on
  the scaling $R$.
  As an application of the previous ideas, we study the detection 
  of point sources characterised by a Gaussian profile  
  $\tau (x) = \exp ({- x^2/2R^2})$, $x = |\vec{x}|$, and Fourier 
  transform  $\tau(q) = R^2 \exp (-(qR)^2/2)$. This is the case we find in 
  CMB experiments, where the profile of the point source is given 
  by the instrumental beam, that can be approximated by a Gaussian.
  
  This profile introduces in a natural way the scale of the source $R$, 
  the scale at which we filter. However, previous works in 1D using the 
  MHW, MF, SAF and the BSAF have shown that the use of a modified scale 
  $\alpha R$ can significantly improve the number of detections 
  (Cay\'on \etal 2000, Vielva \etal 2001a,b, L\'opez-Caniego \etal 2004a,b). 
  Therefore, we generalise the functional form of these filters to 2D and 
  allow for this additional degree of freedom $\alpha$.

  \subsection{The matched filter (MF)} \label{sec:MF}

    Let us introduce a circularly-symmetric filter
    $\Psi (\vec{x};R,\vec{b})$. The filtered field is given by
    equation (\ref{eq:filtered_field}). 
    Now, we express the conditions to obtain a filter for the detection of the 
    source $s(x)=A \tau(x)$ at the origin taking 
    into account the fact that the
    source is characterized by a single scale $R_o$. We assume the
    following conditions: 
    \begin{enumerate}
    \item $\langle w(R_o, \vec{0})\rangle =
      s(0) \equiv A$, i.e. $w(R_o, \vec{0})$ is an \emph{unbiased}
      estimator of the amplitude of the source.
    \item The variance of
      $w(R, \vec{b})$ has a minimum at the scale $R_o$, i.e. it is an
      \emph{efficient} estimator.
    \end{enumerate}
    \noindent
    Then, the 2D filter satisfying these
    conditions is 
    the so-called 
    \emph{matched filter}. 
    As mentioned before, we will allow the filter scale to
    be modified by a factor $\alpha$. If $\alpha=1$ 
    we have the well-known standard 
    matched filter use in the literature.
    For a source with a Gaussian profile, 
    a scale-free power spectrum $P(q)\propto q^{-\gamma}$ and 
    allowing the filter scale to vary 
    through the $\alpha$ parameter, the modified matched filter is
    \begin{equation}\label{eq:mf}
      \psi_{MF}(q) = N(\alpha)z^{\gamma}e^{- \frac{1}{2}z^2}, z\equiv
      q\alpha R, \nonumber 
    \end{equation}
    \noindent where 
    \begin{equation} \label{eq:def_m}
      m\equiv \frac{2 + \gamma}{2}, \ \  
      N(\alpha) = \frac{\alpha^2}{\Delta^m}
      \frac{1}{\pi}\frac{1}{\Gamma(m)}, \ \ 
      \Delta= \frac{2\alpha^2}{(1+\alpha^2)}, 
    \end{equation}
    \noindent
    and $\Gamma$ is the standard Gamma function. 
    The parameters of the filtered background and source are
    \begin{equation}
      \rho(\alpha)=\rho=\sqrt{\frac{m}{1+m}}, \ \ \ \theta_m(\alpha)=
      \alpha R \sqrt{\frac{2}{1+m}}, \ \ \ y_s(\alpha)= \rho \Delta 
    \end{equation}
    
    The corresponding threshold as compared to the standard matched
    filter ($\alpha = 1$) is 
    \begin{equation}
      \frac{\nu(\alpha)}{\nu_{MF(\alpha=1)}}=\alpha^{t-2}\Delta^m,
    \end{equation}
    \noindent
    where
    \begin{equation}  \label{eq:def_t}
      t\equiv \frac{2 - \gamma}{2}.
    \end{equation}

    We remark that for the standard matched filter the curvature does
    not affect the region of acceptance and the linear detector
    $\varphi (\nu ,\kappa )$ is reduced to $\varphi=\nu $. 

  \subsection{The scale-adaptive filter (SAF)}

  \renewcommand{\labelenumi}{3.}

    The scale-adaptive filter (or optimal pseudo-filter) has been proposed by 
    Sanz \etal (2001). 
    The filter is obtained by imposing an additional condition to the
    conditions that define the MF:
    \begin{enumerate}
    \item $w(R, \vec{0})$ has a maximum at $(R_o, \vec{0})$. 
    \end{enumerate}
    Assuming a scale-free power spectrum, $P(q)\propto q^{- \gamma}$,
    a modified scale $\alpha R$ and a Gaussian profile for the source,
    the functional form of the filter in 2D is 
    \begin{equation}\label{eq:saf} 
      \psi_{SAF}(q) = N(\alpha)z^{\gamma}e^{-
      \frac{1}{2}z^2}\left[\gamma + \frac{2t}{m}z^2\right], \ \
      z\equiv q\alpha R, \nonumber 
    \end{equation}  
    \noindent where
    \begin{equation}
      N(\alpha) =
      \frac{\alpha^2}{\Delta^m} \frac{1}{\pi \Gamma(m)}\frac{1}{\gamma
      +\frac{2t}{m}\Delta} 
    \end{equation} 
    \noindent and where $m$ and $\Delta$ are defined as in equation
    (\ref{eq:def_m}), $t$ is defined as in equation (\ref{eq:def_t}).
    The parameters of the filtered background and source are
    \begin{equation}
      \rho(\alpha)=\rho=\sqrt{\frac{m}{1+m}}
      \frac{H_1}{\sqrt{H_2\,H_3}}, \ \ \  \theta_m(\alpha)= \alpha R
      \sqrt{\frac{2}{1+m}} \sqrt{\frac{H_1}{H_3}}, \nonumber 
    \end{equation} 
    \begin{equation}
      y_s(\alpha ) = \sqrt{\frac{m}{1+m}}\sqrt{\frac{H_2}{H_3}} \Delta
      \frac{\gamma + c(1+m)\Delta}{\gamma + cm\Delta}, 
    \end{equation}
    \noindent where $c=2t/m$ and
    \begin{eqnarray}
      H_1 & = & \gamma^2 + 2\gamma c(1+m) + c^2(1+m)(2+m),  \nonumber \\
      H_2 & = & \gamma^2 + 2\gamma cm + c^2m(1+m), \nonumber \\
      H_3 & = & \gamma^2 + 2\gamma c(2+m) + c^2(2+m)(3+m).
    \end{eqnarray}
    
    The corresponding threshold as compared to the standard matched
    filter ($\alpha = 1$) is 
    \begin{equation}
      \frac{\nu}{\nu_{MF(\alpha=1)}} =
      \frac{\alpha^{t-2}\Delta^m
      (\gamma+cm\Delta)}{\sqrt{H_2}}. 
    \end{equation} 

  \subsection{The Mexican Hat wavelet (MH)}

    The MH is defined to be proportional to the Laplacian of the
    Gaussian function in 2D real space  
    \begin{equation}
      \psi_{MH} (x) \propto (1 - x^2 ) e^{-\frac{1}{2}x^2}, \ \ x
      \equiv |\vec{x}|.
    \end{equation}
    Thus, in Fourier space we get the modified Mexican Hat wavelet
    introducing the $\alpha$ parameter as follows 
    \begin{equation}
      \psi_{MH}(q)  = N(\alpha)z^2e^{- \frac{1}{2}z^2},\ \ \ z\equiv
      q\alpha R, \nonumber 
    \end{equation}
    \begin{equation}
      N(\alpha)= \frac{1}{\pi} \left(\frac{\alpha}{\Delta} \right)^2
    \end{equation}
    \noindent Thus, the filtered background and source parameters are
    \begin{equation}
      \rho(\alpha)=\rho=\sqrt{\frac{2+t}{3+t}}, \ \ \
      \theta_m(\alpha)= \alpha R \sqrt{\frac{2}{3+t}}, \ \ \
      y_s(\alpha)=\frac{2}{\sqrt{(2+t)(3+t)}}\Delta, 
    \end{equation}
    \noindent where $m$ and $\Delta$ are defined as in equation
    (\ref{eq:def_m}) and $t$ is defined as in equation (\ref{eq:def_t}). 
    The corresponding threshold is 
    \begin{equation}
      \frac{\nu(\alpha)}{\nu_{MF(\alpha=1)}} = 
      \frac{\alpha^{t-2}\Delta^2}{\sqrt{\Gamma(m)
      \Gamma(2+t)}}.
    \end{equation}
    
  \subsection{The biparametric scale adaptive filter (BSAF)}

  \renewcommand{\labelenumi}{3.}
    
    L\'opez-Caniego \etal (2004b) have shown that removing condition 
    3 defining the SAF and introducing instead the condition
    \begin{enumerate}
    \item $w(R_o,\vec{b})$ has a maximum at $(R_o,\vec{0})$,
    \end{enumerate}
    leads to the new filter
    \begin{equation}
      \psi (q) \propto \frac{\tau(q)}{P(q)}\left[1+c(qR)^2\right],
      \label{eq:eqnndf}
    \end{equation}
    \noindent where $c$ is an arbitrary constant related to the
    curvature of the maximum. For the case of a scale-free spectrum
    and allowing for a modified scale $\alpha R$, the filter is given
    by the parameterized equation 
    \begin{equation}\label{eq:bsaf}
      \psi_{BSAF}(q) = N(\alpha)z^{\gamma}e^{- \frac{1}{2}z^2}\left(1+
      cz^2\right), \ \ z\equiv q\alpha R,\nonumber 
    \end{equation}
    \begin{equation}
      N(\alpha) = \frac{\alpha^2}{\Delta^m}
      \frac{1}{\pi}\frac{1}{\Gamma(m)}\frac{1}{1 +cm\Delta}, 
    \end{equation} 
    \noindent where
    $m$ and $\Delta$ are defined as in equation
    (\ref{eq:def_m}).
    We remark that $c = 0$ leads to the MF and  if $c\equiv
    2t/m \gamma$, with $t$ defined as in equation (\ref{eq:def_t}), 
    the BSAF defaults to the SAF. The parameters
    of the filtered background and source are
    \begin{equation}
      \rho(\alpha)=\rho=\sqrt{\frac{m}{1+m}}
      \frac{D_1}{\sqrt{D_2\,D_3}}, \ \ \ \theta_m(\alpha)= \alpha R
      \sqrt{\frac{2}{1+m}} \sqrt{\frac{D_1}{D_3}},\nonumber 
    \end{equation}
    \begin{equation}
      y_s(\alpha)=\sqrt{\frac{m}{1+m}}\sqrt{\frac{D_2}{D_3}} \Delta
      \frac{1 + c(1+m)\Delta}{1 + cm\Delta}, 
    \end{equation}
    \noindent where 
    \begin{eqnarray}
      D_1 & = & 1 + 2c(1+m) + c^2(1+m)(2+m),\nonumber \\
      D_2 & = & 1 + 2cm + c^2m(1+m),\nonumber \\
      D_3 & = & 1 + 2c(2+m) + c^2(2+m)(3+m).
    \end{eqnarray}
    
    The equivalent threshold is given by
    \begin{equation}
      \frac{\nu(\alpha)}{\nu_{MF(\alpha=1)}} = 
      \frac{\alpha^{t-2}\Delta^m
      (\gamma+cm\Delta)}{\sqrt{D_3}}. 
    \end{equation}
    
\section{ANALYTICAL RESULTS} \label{sec:numerical}

In this section we will compare the performance of the different
filters previously introduced.
We use as example the
interesting case of
white noise as background. This is a fair approximation to the
case presented in 
Figure~\ref{fig:example_residual}, where the sources are embedded in a
background that is a combination of instrumental noise (approximately
Gaussian) and a small contribution of residual foregrounds that have
not been perfectly separated.
For
this example, we will consider sources with intensities distributed
uniformly between zero and a upper cut-off. 

The comparison of the filters is performed as follows: we fix the
number density of spurious detections, the same for all the
filters. Then, for any given filter we calculate the quantities
$\sigma_n$, $\rho$ and $y_s$. Using equation (\ref{eq:nbstar1}) it is
possible to calculate the value of $\varphi_*$ that defines the
region of acceptance. Then we calculate the number density of real
detections using equation (\ref{eq:nstar1}). The filter that leads to
the highest number density of detections will be the preferred one. We do this
for different values of $\alpha$ in order to test how the variation of
the filtering scale affects the number of detections.

\subsection{\emph{A priori} probability distribution}

  As mentioned before, we will test a \pdf 
  of source intensities that is uniform in the interval $0 \leq A
  \leq A_c$. In terms of normalized intensities, we have the \pdf   
  \begin{equation}
    p(\nu_s ) = \frac{1}{\nu_c}, \ \ \ \nu_s\in [0, \nu_c]. 
  \end{equation}
  \noindent
  We will consider a cut-off in the amplitude of the sources 
  such that $\nu_c=2$ after filtering with he standard MF, that is, we
  will focus on the case of faint sources that would be very difficult
  to detect if no filtering was applied. Note
  that while the value $\nu_c$ is different for each filter (because
  each filter leads to a different dispersion $\sigma_0$ of the
  filtered field), 
  the distribution in source intensities $A$ is the same for all the
  cases.

\subsection{Results for white noise}

  We want to find the optimal filter in the sense of maximum number 
  of detections. For the sources, we use a uniform distribution with 
  amplitudes in the interval $A\in[0,2]\sigma_0$, where $\sigma_0$ is 
  the dispersion
  of the linearly-filtered 
  map with the standard MF. We focus on the interesting case of white 
  noise ($\gamma = 0$) and explore different values of $n^*_b$ and $R$. 

\begin{figure}
\epsfxsize=75mm
\includegraphics[width=14cm]{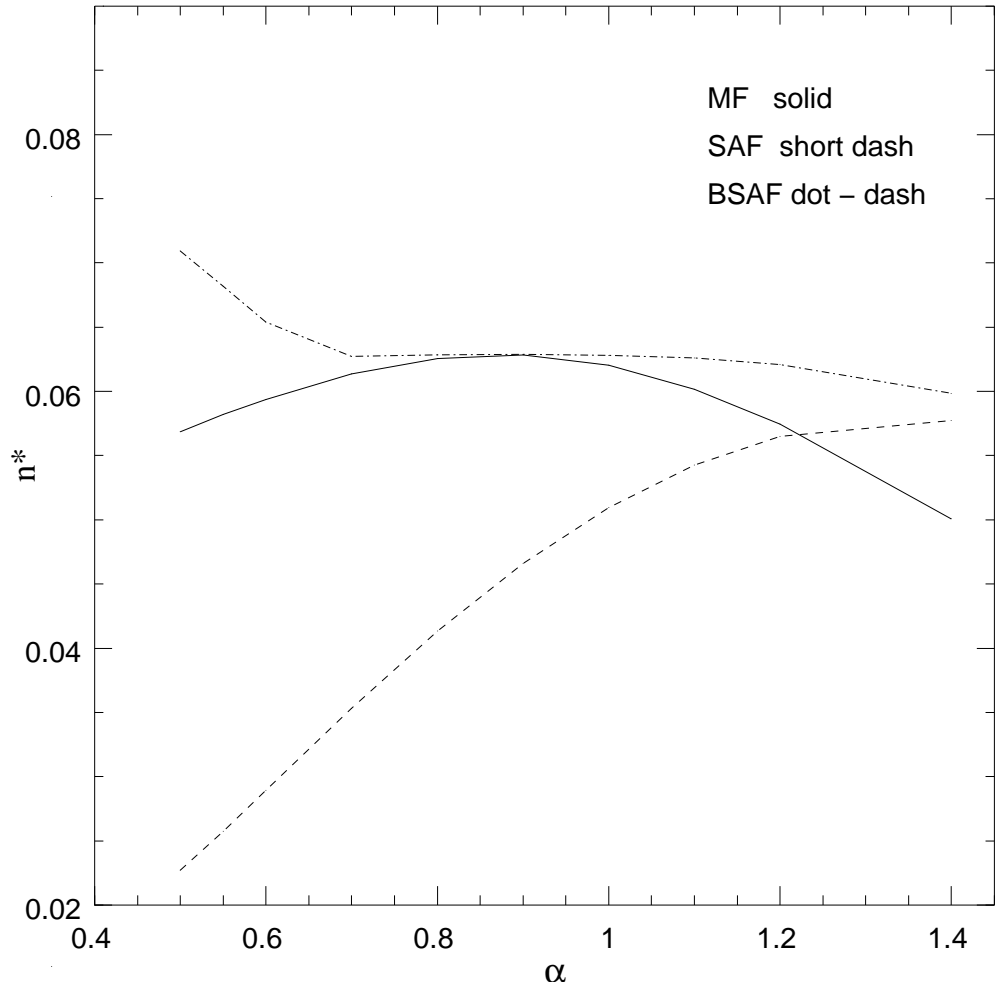}
\caption{The expected number density of detections $n^*$ as a 
function of  $\alpha$ for $\gamma=0$ for the BSAF (c has 
been obtained by maximising the number of detections for each 
value of $\alpha$), MF and SAF filters. We consider the case 
$R=1.5$, $n^*_b=0.01$}
\label{fig:r15_01}
\end{figure}

\begin{figure}
\epsfxsize=75mm
\includegraphics[width=14cm]{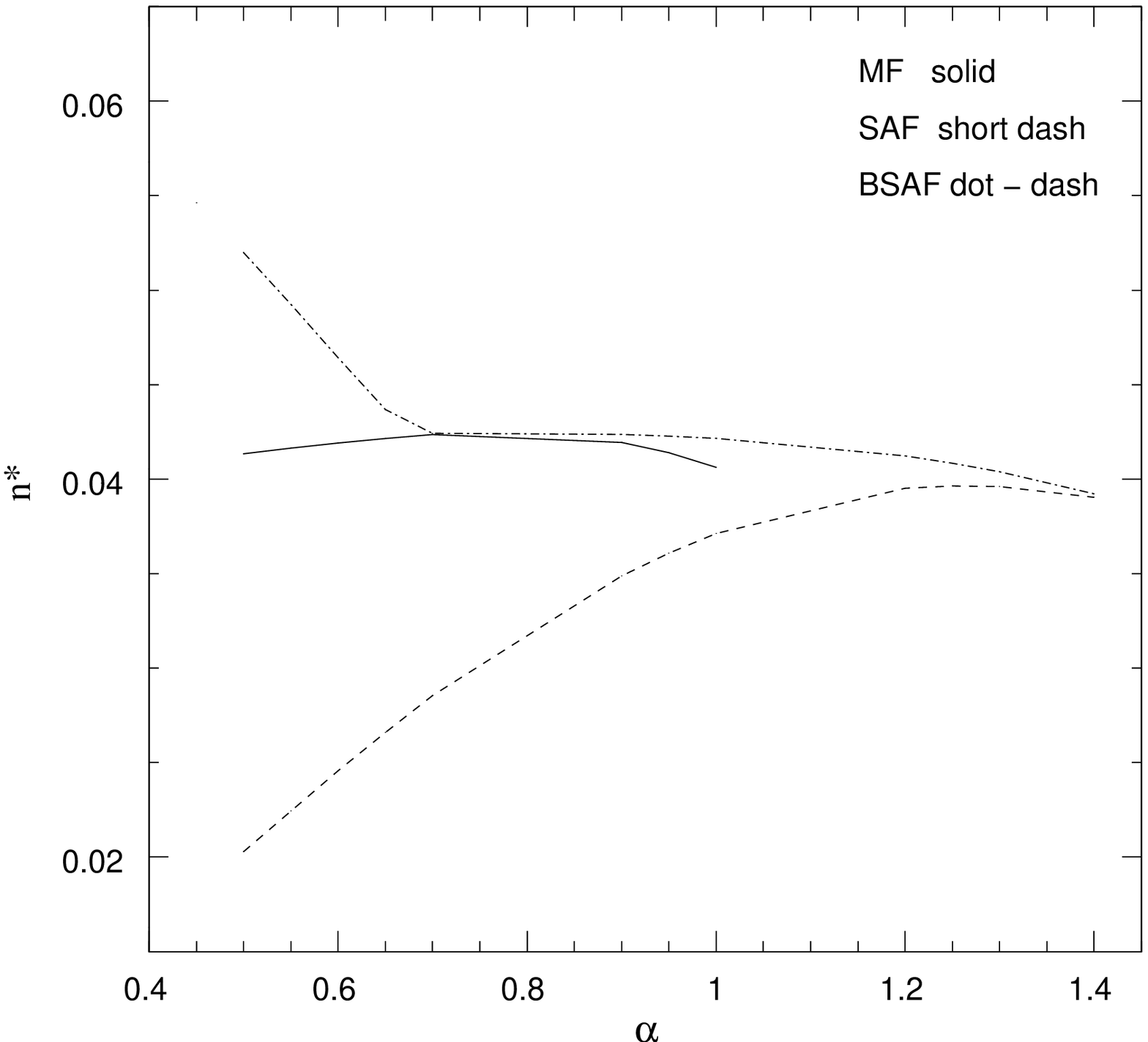}
\caption{The expected number density of detections $n^*$ as a 
function of  $\alpha$ for $\gamma=0$ for the BSAF (c has been obtained 
by maximising the number of detections for each value of $\alpha$), 
MF and SAF filters. We consider the case $R=2$, $n^*_b=0.01$}
\label{fig:r2_01}
\end{figure}

  We study the performance of the different filters as a function of
  $\alpha$. This allows us to test how the variation of the natural
  scale of the filters helps the detection. In the case of the BSAF,
  which has an additional free parameter, $c$ in equation
  (\ref{eq:bsaf}), for each value of $\alpha$ we determine numerically
  the value
  of $c$ that gives the highest number of detections. Then the BSAF
  with such $c$ parameter (that is a function of $\alpha$, $n_b^*$ and
  $R$) is compared with the other filters. 
  
  In Figure \ref{fig:r15_01}, we plot the expected number density of 
  detections $n^*$ for different values of $\alpha$,  $R=1.5$ pixels 
  and $n^*_b=0.01$. Note that for the 2D case the MHW and SAF are the 
  same filter for $\gamma =0$ and we have only included the latter in 
  our figures. In this case, the curve for the BSAF always goes above 
  the other filters. The maximum number of detections is found for
  small values of  $\alpha$. 
  In this region, the improvement of the BSAF with respect to the standard 
  matched filter is of the order $\simeq 15 \%$. 

  In Figure \ref{fig:r2_01}, 
  we show the results for $R=2$. We have increased the beam width as 
  compared with the previous example and leave unchanged the number density 
  of false detections. The BSAF outperforms all the other filters, and for 
  small values of $\alpha$ the improvement is of the order  $\simeq 40 \%$. 
  Note that in this figure the MF takes values $\alpha \in[0,1]$. For 
  greater values of $\alpha$, with $R=2$ and $n^*_b=0.01$, we can not 
  solve for $\varphi_*$ in the implicit equation (\ref{eq:nbstar1}) and 
  can not calculate $n^*$. 
  
  We remark that filtering at scales much smaller than the scale of the pixel 
  does not make sense. This is due to the fact that we are not including the 
  effect of the pixel in our theoretical calculations and, thus, the results 
  would not exactly follow what would be found in a real image. Therefore, 
  we present the results only for those values of $\alpha$ such that
  $\alpha R$ is at least $\sim 1$.

  \begin{table}
    \begin{center}
      \label{ref:tabla1_n}
      \begin{tabular}{ccccccc}
	\hline
	$R$ &  $n^*_b$ & $\alpha$ & c &$n^*_{BSAF}$ & $n^*_{MF}$ & $RD [\%]$ \\
	\hline
	\hline
	1.5 & 0.005 &  0.5 & -0.44 & 0.0507 & 0.0484  & 4.7 \\
	& 0.01  &  0.5 & -0.46 & 0.0709 & 0.0620 & 14.3 \\
	\hline
	2 & 0.005 &  0.4  & -0.54  & 0.0396 & 0.0335 & 18.2  \\
	& 0.01  &  0.4  & -0.54  & 0.0567 & 0.0406 & 39.6  \\
	\hline
	2.5 & 0.005  & 0.3 & -0.64 & 0.0320 & 0.0245  & 33.3 \\
	\hline
      \end{tabular}
      \caption{Number density of detections $n^*$ for the BSAF and 
	the standard MF($\alpha=1$) with optimal values of c and 
	$\alpha$ for different values of  $n^*_b$ and $R$. RD means 
	relative difference in number densities in percentage: 
	$RD\equiv 100 (-1 + n^*_{BSAF}/n^*_{MF})$.}
    \end{center}
  \end{table}
   
\section{CONCLUSIONS} \label{sec:conclusions}

Several techniques have been introduced in the literature to detect 
point sources in two-dimensional images. Examples of point sources in astronomy 
are far galaxies as detected by CMB experiments. An approach that has been thoroughly used in
the literature for this case consists in linear filtering the data and
applying 
detectors based on thresholding. Such approach uses only information
on the amplitude of the sources: the potentially useful information
contained in the local spatial structure of the peaks is not used at
all. 
In 
our work, we design a detector
based on peak statistics that uses 
the information contained in the amplitude, curvature and 
shear of the maxima. These quantities describe the local properties of
the maxima and are used to distinguish statistically between peaks due
to background fluctuations and peaks due to the presence of a source.

We derive a Neyman-Pearson detector (NPD) that considers number densities of
peaks 
which
leads 
to a sufficient detector that, in the case of the
spherically symmetric sources that we consider, is linear in the amplitude
and curvature of the sources. For this particular case, then, the
information of the shear of the peaks is not relevant. In
other cases, however, it could be useful.

It is a common practice in astronomy to linear filter the images in
order to enhance very faint point sources and help the detection. The
best filter would be the one 
that makes easier to distinguish between peaks coming from the
background alone and those due to the presence of a source, according
to the information used by the detector. In
the case of simple thresholding, that considers only the amplitude of
the peaks, the answer to the question of which is the best filter (in
the previous sense) is well known: the standard matched filter. But
in the case of the Neyman-Pearson detector, that considers other
things apart from mere amplitudes, this is not longer true. 

We have 
compared three commonly used filters in the literature
in order to assess which one of them performs better when detecting
sources with
our scheme. In addition, 
we have designed a filter such that 
optimizes the number of true detections for a fixed number of 
spurious sources. 
The optimization of the number of true detections is performed
by using the \emph{a priori} \pdf \ of the amplitudes of the sources.
This filter depends on two free parameters and it is therefore 
called biparametric scale adaptive filter (BSAF). 
By construction, the functional form of the BSAF includes the standard
MF as a particular case and its performance in terms of number of true
detections for a fixed number of spurious detections must be at least
as good as the standard MF's one.

Following the work
done in the 1D case, we 
generalize the functional form of the filters to 2D and introduce 
an extra degree of freedom $\alpha$ that will allow us to filter 
at different scales  $\alpha R$, where $R$ is the scale of the source. 
This significantly improves the results.

We have considered an interesting case, a uniform distribution of weak 
sources with amplitudes $A\in[0,2]\sigma_0$, where $\sigma_0$ is 
the dispersion of the field filtered with the standard matched filter, 
embedded in white noise ($\gamma=0$). We have tested different values
of the source size $R$ and of the number density of spurious detections
$n^*_b$
that
we fix. We find that the BSAF improves the number density of 
detections up to $\simeq 40 \%$ with respect to the standard 
MF ($\alpha=1$) for certain cases. Note that since the
Neyman-Pearson detector for the standard MF ($\alpha=1$) defaults
to the classic thresholding detector that is commonly used in
astronomy, the results of this work imply that it is possible, under
certain circumstances, to    
detect more point sources than in the classical approach. 

The generalization of these ideas to other source profiles and 
non-Gaussian backgrounds is relevant and will be discussed in a future work.

\section*{Acknowledgments}

The authors thank Patricio Vielva for useful discussions.
MLC thanks the Ministerio de Ciencia y Tecnolog\'\i a (MCYT) 
for a predoctoral FPI fellowship. RBB thanks the MCYT and 
the Universidad de Cantabria for a Ram\'on y Cajal contract.
DH acknowledges support from the European Community's Human 
Potential Programme under contract HPRN-CT-2000-00124, CMBNET,
and from a ISTI fellowship since September 2004.   
We acknowledge partial support from the Spanish MCYT project 
ESP2002-04141-C03-01 and from the EU Research Training Network 
`Cosmic Microwave Background in Europe for Theory and Data Analysis'.

\appendix

  \section{Appendix}

  We will show in this Appendix that $\varphi (\nu, \kappa) \geq \varphi_*$ 
  given in equation (\ref{eq:toapp}) is a sufficient linear detector, i.e., the detector is linear 
  in the threshold $\nu$ and the curvature $\kappa$ and the data it uses is a sufficient statistic to decide
  if a peak is a source (independent of the {\it a priori} probability $P(\nu_s)$). The ratio $L(\nu ,\kappa, \epsilon |\nu_s) \equiv  n(\nu ,\kappa,
  \epsilon |\nu_s)/n_b(\nu ,\kappa, \epsilon )$ can be explicitly
  written as
  \begin{equation}
    L(\nu ,\kappa, \epsilon |\nu_s) = e^{\varphi \nu_s - \frac{1}{2}
      (\nu_s^2 + (\rho \nu_s - 2\kappa_s)^2)} 
  \end{equation}
  The criterion for detection can be written as 
  \begin{equation}
    \L(\nu ,\kappa )\equiv  \int_0^{\infty} d\nu_s\, p(\nu_s)  L(\nu
    ,\kappa |\nu_s) \geq L_*, 
  \end{equation}
  where $L_*$ is a constant. $L$ is a function of $\varphi$,
  \begin{equation}\label{eq:app_3}
    \varphi (\nu ,\kappa )\equiv a\nu + b \kappa, ~ a=\frac{1 - \rho
      y_s}{1 - \rho^2}, ~ b=\frac{y_s - \rho }{1 - \rho^2}. 
  \end{equation}
  By differentiating $L$ with respect to $\varphi$ we find that
  \begin{equation}
    \frac{\partial L}{\partial \varphi } = 
    \int_0^{\infty} d\nu_s\, p(\nu_s)\nu_s e^{\varphi \nu_s -
      \frac{1}{2}(\nu_s^2 +(\rho \nu_s - 2\kappa_s)^2)} \geq 0 ,
  \end{equation}
  \noindent and therefore setting a threshold in $L$ is equivalent
  to setting a threshold in $\varphi$:
  \begin{equation}
    \L(\nu,\kappa)\geq L_* \Leftrightarrow  \varphi (\nu, \kappa) 
    \geq \varphi_*,
  \end{equation}
  \noindent where $\varphi (\nu ,\kappa )$ is given by equation
  (\ref{eq:app_3}) and $\varphi_*$ is a constant.


\begin{thebibliography}{}

\bibitem{} Baccigalupi, C., Bedini, L., Burigana, C., de Zotti, G.,
  Farusi, A., Maino, D., Maris, M., Perrota, F., Salerno, E.,
  Toffolatti, L. \& Tonazzini, A., 2000, ``Neural networks and
  separation of Cosmic Microwave Background and astrophysical signals
  in sky maps'', MNRAS, 318, 769.

\bibitem{} Barreiro, R. B., Sanz, J. L., Mart\'\i nez-Gonz\'alez,
  Cay\'on, L., Silk, J., 1997, ``Peaks in the Cosmic Microwave
  Background: flat versus open models'', ApJ, 478, 1B

\bibitem{} Barreiro, R.B., Hobson, M.P., Banday, A.J., Lasenby, A.N.,
  Stolyarov, V., Vielva, P., G\'orski, K.M., 2004, ``Foreground
  separation using a flexible maximum-entropy algorithm: an
  application to COBE data'', MNRAS, 351, 515.

\bibitem{} Bedini, L., Herranz, D., Salerno, E., Baccigalupi, C.,
  Kuruo\u glu, E.E., Tonazzini, A., 2004, ``Separation of correlated
  astrophysical sources using multiple-lag covariance matrices'',
  submitted to EURASIP JASP in the Special Issue ``Applications of
  Signal Processing in Astrophysics and Cosmology".

\bibitem{} Bond, J. R., Efstathiou, G., 1987, ``The statistics of
  cosmic background radiation fluctuations'', MNRAS, 226, 655.

\bibitem{} Bouchet, F.R. \& Gispert, R., 1999, `` Foregrounds and CMB
  experiments I. Semi-analytical estimates of contamination'', New
  Astronomy, vol. 4, no. 6, 443. 

\bibitem{} Cay\'on, L., Sanz, J. L., Barreiro, R. B., Mart\'\i
  nez-Gonz\'alez, E., Vielva, P., Toffolatti, L., Silk, J., Diego,
  J. M. \& Arg\"ueso, F., 2000, ``Isotropic wavelets: a powerful tool
  to extract point sources from cosmic microwave background maps'',
  MNRAS, 757, 761.
 	  
\bibitem{} Chiang, L.Y., J\o rgensen, H.E.,  Naselsky I.P., Naselsky
  P.D.,  Novikov I.D. \& Christensen, P.R., 2002, ``An adaptive filter
  for the PLANCK Compact Source Catalogue construction'', MNRAS, 335,
  1054.

\bibitem{} Delabrouille, J., Cardoso, J.F. \& Patanchon, G., 2003,
  ``Multidetector multicomponent spectral matching and applications
  for Cosmic Microwave Background data analysis'', MNRAS, 346, 1089. 

\bibitem{} Guiderdoni, B., Hivon, E., Bouchet, F.R., Maffei, B., 1998,
  ``Semi-analytic modelling of galaxy evolution in the IR/submm
  range'', MNRAS, 295, 877.

\bibitem{} Herranz, D., Sanz, J.L., Barreiro, R. B., E. Mart\'\i
  nez-Gonz\'alez, 2002, ``Scale-Adaptive filters for the
  detection/separation of compact sources'', ApJ, 580 610-625.

\bibitem{} Hobson, M.P., Jones, A.W., Lasenby, A.N., Bouchet F.R.,
  1998, ``Foreground separation methods for satellite observations of
  the cosmic microwave background'', MNRAS, 300, 1.

\bibitem{} Hobson, M.P., Barreiro, R.B., Toffolatti, L., Lasenby,
  A.N., Sanz, J.L., Jones, A.W., Bouchet, F.R., 1999, ``The effect of
  point sources on satellite observations of the cosmic microwave
  background'', MNRAS, 306, 232.

\bibitem{} L\'opez-Caniego, M., Herranz, D., Barreiro, R. B.,  Sanz,
  J.L., 2004a, ``A Bayesian approach to filter design: detection of
  compact sources'', Proceedings to the SPIE Conference on Electronic
  Imaging, San Jose, USA, January 2004.

\bibitem{} L\'opez-Caniego, M., Sanz, J.L., Herranz, D., Barreiro,
  R. B., 2004b, ``Filter design for the detection of compact sources
  based on the Neyman-Pearson detector'', MNRAS submitted.  

\bibitem{} Maino, D., Farusi, A., Baccigalupi, C., Perrotta, F.,
  Banday, A. J., Bedini, L., Burigana, C., De Zotti, G., G\'orski,
  K. M., Salerno, E, 2002, ``All-sky astrophysical component
  separation with Fast Independent Component Analysis (FASTICA)'',
  MNRAS, 334, 53.

\bibitem{} Rice, S. O., 1954, ``Selected papers on Noise and Stochastic
  Processes", ed. N. Wax, Dover Publ. Inc. (N. Y.).

\bibitem{} Sanz, J. L., Herranz, D. \& Mart\'\i nez-Gonz\'alez, E.,
  2001, ``Optimal detection of sources on a homogeneous and isotropic
  background'', ApJ, 552, 484.

\bibitem{} Stolyarov, V., Hobson, M.P., Ashdown, M.A.J., Lasenby,
  A.N., 2002, ``All-sky component separation for the Planck mission'',
  MNRAS, 336, 97.

 \bibitem{} Tegmark, M. \& de Oliveira-Costa, A., 1998, ``Removing
 Point Sources from Cosmic Microwave Background Maps'', ApJL, 500, 83.
  	
\bibitem{} Toffolatti, L., Arg\"ueso, F., De Zotti, G., Mazzei, P.,
  Franceschini, A., Danese, L. and Burigana, C. 1998, ``Extragalactic
  source counts and contributions to the anisotropies of the cosmic
  microwave background: predictions for the Planck Surveyor mission'',
  MNRAS, 297, 117.
	
\bibitem{} Tucci, M., Mart\'\i nez-Gonz\'alez, E., Toffolatti, L.,
  Gonz\'alez-Nuevo, J., De Zotti, G., 2004, ``Predictions on the
  high-frequency polarization properties of extragalactic radio
  sources and implications for polarization measurements of the cosmic
  microwave background'', MNRAS, 349, 1267.

\bibitem{} Vielva, P., Mart\'\i nez-Gonz\'alez, E., Cay\'on, L.,
  Diego, J. M., Sanz, J. L. \& Toffolatti, L., 2001a, ``Predicted
  Planck extragalactic point-source catalogue'', MNRAS, 326, 181.
  
\bibitem{} Vielva, P., Barreiro, R. B., Hobson, M.P., Mart\'\i
  nez-Gonz\'alez, E., Lasenby, A.N., Sanz, J. L. \& Toffolatti, L.,
  2001b, ``Combining maximum-entropy and the Mexican Hat wavelet to
  reconstruct the microwave sky'', MNRAS, 328, 1.
  
\bibitem{} Vielva, P., Mart\'\i nez-Gonz\'alez, E., Gallegos, J.E.,
  Toffolatti, L., Sanz, J.L., 2003, ``Point source detection using the
  Spherical Mexican Hat Wavelet on simulated all-sky Planck maps'',
  MNRAS, 344, 89.

\end{thebibliography}
\end{document}